\RequirePackage{fix-cm}
\documentclass[twocolumn]{svjour3}          % twocolumn
\usepackage{graphicx}
\usepackage{color}
\usepackage[round]{natbib} 
\usepackage[switch]{lineno}
\journalname{Climate dynamics}
\begin{document}

\title{Record occurrence and record values in daily and monthly temperatures}



\author{G. Wergen \and A. Hense \and J. Krug}


\institute{G. Wergen \and J. Krug \at
              Institute for Theoretical Physics \\
	      University of Cologne, 50939 Germany\\
              \email{gw@thp.uni-koeln.de}           %  \\
           \and
           A. Hense \at
	      Meteorological Institute \\
	      University of Bonn, Auf dem H\"ugel 20 \\ 53121 Bonn \\
              \email{ahense@uni-bonn.de}
}

\date{Received: date / Accepted: date}

\maketitle

\begin{abstract} We analyze the occurrence and the values of record-breaking temperatures in daily and monthly temperature observations. Our aim is to better understand and quantify the statistics of temperature records in the context of global warming. Similar to earlier work we employ a simple mathematical model of independent and identically distributed random variables with a linearly growing expectation value. This model proved to be useful in predicting the increase (decrease) in upper (lower) temperature records in a warming climate. Using both station and re-analysis data from Europe and the United States we further investigate the statistics of temperature records and the validity of this model. The most important new contribution in this article is an analysis of the statistics of record values for our simple model and European reanalysis data. We estimate how much the mean values and the distributions of record temperatures are affected by the large scale warming trend. In this context we consider both the values of records that occur at a certain time and the values of records that have a certain record number in the series of record events. We compare the observational data both to simple analytical computations and numerical simulations. We find that it is more difficult to describe the values of record breaking temperatures within the framework of our linear drift model. Observations from the summer months fit well into the model with Gaussian random variables under the observed linear warming, in the sense that record breaking temperatures are more extreme in the summer. In winter however a significant asymmetry of the daily temperature distribution hides the effect of the slow warming trends. Therefore very extreme cold records are still possible in winter. This effect is even more pronounced if one considers only data from subpolar regions. 

\end{abstract}

\tableofcontents

\section{Introduction} In the context of global warming, record-breaking temperatures have received considerable attention recently. In newspapers and in television one frequently hears of hottest summer days, extreme heat streaks, or record breaking storms. Extreme and record breaking weather events are not only interesting for the observer but can also have a big impact on agriculture, economy and human life. If one considers record-breaking events in climatology, these should of course always be seen in the context of global warming. The crucial question is: How much are the climate records we encounter today affected by the evident climatic change \citep{Solomon2007} of the last decades?

\begin{figure}
\begin{center}\includegraphics[width=0.48\textwidth]{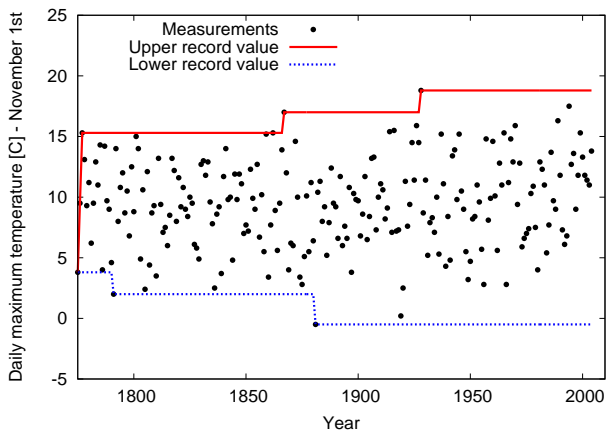}
\caption{\label{Fig:prag} Daily maximum temperature measurements for
  November 1st in Prague for the time span from 1772 to 2010. The figure
  illustrates the progression of the upper and lower record
  values. The \textit{full} red line gives the progression of the highest
  temperature in the time-series and the dotted blue line the progression of the lowest one. In each case the number of steps is the respective upper or lower record number. The statistics of those record progressions, the number and the height of the steps is the subject of this paper. }\end{center}
\end{figure} 

While the study of extremal events in general is very important in climatology (Easterling et al., \citeyear{Easterling1997}; Stott et al., \citeyear{Stott2004}; Brown et al., \citeyear{Brown2008}; Cattiaux et al., \citeyear{Cattiaux2009}; Meehl et al., \citeyear{Meehl2007}; Min et al., \citeyear{Minetal2009}), record temperatures have not received much attention until recently. Even though it is intuitively clear that increasing temperatures should result in a higher than average number of hot day records, this effect was not studied and detected in observational data for a long time. However, there has been some research on the statistics of temperature records in the last years. Redner and Peterson \citeyearpar{Redner2006} analyzed the statistics of record temperatures for daily temperature measurements from Philadelphia over a 126 year time-span. Due to the fact that they only used data from a single station, their analysis did not establish a significant connection between the increase of global mean temperatures and local record-breaking events, but made important theoretical contributions to the subject. There is also some earlier work by Benestad \citeyearpar{Benestad2003}, who analyzed the occurrence of records for stations in Scandinavia.

Meehl et al \citeyearpar{Meehl2009} found a significant effect of slowly changing temperatures on the occurrence of records for daily temperatures at weather stations in the United States. In particular they demonstrated that in an analysis starting from 1970 the rate at which upper records occurred declined more slowly than the rate of lower records. In 2010 two of us obtained similar results in an independent study \citep{Wergen2010}. We performed an extensive analysis of European and American station data and combined it with a simple mathematical model \citep{Franke2010} to quantify the effect of slow temperature changes upon the records of daily temperatures. For the European station data covering the time period 1976-2005, we found that on average 5 of the 17 high temperature records recorded at one station in one year can be attributed to the observed slow increase of temperature. 

Our findings and our analytical model were discussed and confirmed by Elguindi et al \citeyearpar{Elguindi2012} using gridded data from regional climate models. They also made predictions for the spatial distribution of record temperatures in Europe for the future based on model data from the ENSEMBLES project. Newman et al \citeyearpar{Newman2010} analyzed record breaking temperatures at a very high resolution from the Mauna Loa Observatory on Hawaii for the time-span from 1977 to 2006. They also presented evidence for slowly increasing temperatures in the occurrence of records. However, in their data they found that while the rate of cold records is significantly decreased, the number of hot records remained unchanged. Rahmstorf and Coumou \citeyearpar{Rahmstorf2011} considered monthly mean temperatures from a weather station in Moscow and could show that the number of hot records in these mean values increased significantly. They also discussed the effect of climatic change on the occurrence of global-mean temperature records based to a large extent on numerical experiments. Recently, Coumou et al. \citeyearpar{Coumou2013} studied the occurrence of records in monthly mean values for a worldwide gridded data set and found a fivefold increase in the number of upper records. In this context, they could confirm the validity of our simple analytical model for the monthly averages.

The purpose of this paper is to provide a detailed analysis of the statistical properties of record-breaking temperatures also from a theoretical point of view. The main idea behind this analysis is illustrated in Fig. \ref{Fig:prag}. We consider the progression of the records and record values in time series of temperature measurements for individual calendar days and months. This way, the daily measurement are always one year apart from each other and, to a good approximation, their statistics can therefore be compared to uncorrelated random variables. We use both station and gridded re-analysis data of daily and monthly mean and maximum temperatures to get a more complete picture in space and time. 

In particular we will consider monthly mean temperatures at single stations for the continental United States and gridded temperature data for Europe. In 2010 we already considered daily station data from the United States and had difficulties to quantify the effect of slow changes in temperature with our model \citep[see][]{Wergen2010}. In this paper we will find that if one analyses monthly mean values, both the increase of the upper record number and the decrease of the lower record number is much more pronounced than in the daily data. The reason for this is the strongly reduced level of variability in the monthly averages \citep[see][]{Coumou2013}. For the European daily data we already found a strong effect of slowly increasing temperatures, since the increase of the mean temperature was stronger and the standard deviation was smaller in this data set. Therefore and because of the high density and homogeneity of the gridded temperature data \citep{EOBS} we decided to analyze the statistics of record values of these European data.

At this point we will give a brief outline of the article: Before we
introduce and discuss the different data sets, we will give an
overview and some new results on the record statistics of independent
and identically distributed (iid) random variables (RV's) with a
linear drift. This Linear Drift Model (LDM) discussed by
\cite{Franke2010} is very important for our understanding of
temperature records.  The results for the occurrence of records in the
LDM have been published before, but we briefly describe them to make the article self contained. %We discuss the statistical effects of the fact that the daily temperature recordings are discrete measurements that allow for records to get tied. This discreteness effect increases or decreases the occurrence of record temperatures depending on whether or not we count those ties. We give a new and simple mathematical model that describes the discreteness effects. 

In section 3 we introduce the data sets that were mentioned above. We describe the statistical properties of the measurements and their time-dependence and discuss how well the observational data fit the LDM. For that purpose we analyzed the time-dependence of the mean and the standard deviation of the recordings. In the following section 4 we will then first consider the occurrence of records in the different data sets, in particular the record rate in the daily and monthly temperature recordings. We analyzed the record rate in the United States with respect to the different seasons to find out when the effect of slowly evolving temperatures on the record statistics is strongest. In particular we will discuss the ratio between the number of upper records and the number of lower records and compare it to the predictions from our analytical model. 

Section 5 is again about theoretical aspects. Here we will discuss the statistics of record values within the LDM using both an analytical approach and numerical simulations. We quantify the effect of the linear drift on the mean value of record values that occur at a certain time step and of record values that have a certain record number in the series of record events. The aim is to understand if the record events we have to expect in the presence of slowly increasing temperatures are more extreme or more variable than without any climatic change. 

In the subsequent section 6 we will then compare the findings of section 5 to the observational data we already introduced and discussed in sections 3 and 4. In section 6 we will consider the behaviour of the mean values of record events as well as their full distributions. For that purpose we will introduce a simple rescaling of the observational data to account for seasonal and spatial variations in the standard deviation of the time series. Then we will discuss both the values of records that occur in a given year as well as the values of records that have a certain record number. 

Finally in section 7 we summarize and evaluate our findings and discuss them in the context of ongoing research in the field of temperature records.

\section{Theory: Record occurrence in the presence of linear drift}
\subsection{Record statistics of iid. RV's}
We consider time series of uncorrelated random variables (RV) $X_k$ from continuous probability densities $f_{X_k}\left(x_k\right), k \in \{1,2\ldots,n\}$. As mentioned earlier, an upper (lower) record in the $n$th step is an entry $X_n$ that is larger (smaller) than all previous entries $X_k$ with $k<n$. The basic properties of record events in such time series can for instance be found in \citep{Arnold1998,Glick1978,Nevzorov2001}. In the special case of identically and independently distributed (iid) RV's from a single probability density $f_X\left(x\right)$ the probability that the $n$th entry is a record is simply given by $P_n = 1/n$ (cf. \citep{Arnold1998,Nevzorov2001}). This holds, because of the symmetry of the problem, both for upper and lower records. From now on we will call the probability $P_n$ that the $n$th entry in the series is a record the record rate. In the iid case the mean number of records $R_n$ up to the $n$th step can be obtained by summing over the record rate and for large $n$ we 
find:
\begin{eqnarray}
 R_n = \sum_{k=1}^n P_k = \sum_{k=1}^n \frac{1}{k} \approx \textrm{ln}\left(n\right) + \gamma.
\end{eqnarray}
Here, $\gamma \approx 0.577 215...$ is the Euler-Mascheroni constant \citep{Arnold1998,Glick1978}. In this case, one can prove that record events are uncorrelated (Sibani and Littlewood, \citeyear{Wergen2011}) and, if one goes to a logarithmic time scale one finds that they form a Poisson process \citep{Sibani1993}. Another important feature of this result for records of iid RV is that $P_n$ and $R_n$ are completely independent of the shape of the underlying probability density $f_X\left(x\right)$.

\subsection{Linear Drift}
In general, when the RV's $X_k$ are not identically distributed, it is more difficult to compute the record rate $P_n$. For independent RV's $X_k$ from a series of arbitrary continuous distributions $f_{X_k}\left(x_k\right)$ the upper record rate is given by the following integral \citep{Arnold1998,Glick1978}:
\begin{eqnarray}\label{pngeneral}
 P_n = \int \mathrm{d}x_n\;f_{X_n}\left(x_n\right) \prod_{k=1}^{n-1} F_{X_k}\left(x_n\right)
\end{eqnarray}
where $F_{X_k}\left(x_n\right)$ is the probability distribution function of $f_{X_k}\left(x_k\right)$ with $F_{X_k}\left(x_n\right) = \int^{x_n} \mathrm{d}x\;f_{X_k}\left(x\right)$. Here, the integrand is just the probability that the $n$th entry has a value of $x_n$ times the probability that all previous entries are smaller, which is represented by the product. This probability is then integrated over all possible values for a record in the $n$th step $x_n$. Analogous to this the lower record rate $P_n^{\star}$ is given by:
\begin{eqnarray}
 P_n^{\star} = \int \mathrm{d}x_n\; f_{X_n}\left(x_n\right) \prod_{k=1}^{n-1} \left(1-F_{X_k}\left(x_n\right)\right).
\end{eqnarray}
The LDM was first considered by Ballerini and Resnick \citeyearpar{Ballerini1985,Ballerini1987} and later also by Borovkov \citeyearpar{Borovkov1999}. In this model we consider iid RV's $Y_k$ with a linear drift of the following form:
\begin{eqnarray}
 X_k = Y_k + ck,
\end{eqnarray}
with a constant $c$. In this case $f_{X_k}(x)$ is simply given by $f_{X_k}(x) = f(x-ck)$ with fixed $f\left(x\right)$. The underlying distributions have all the same shape, but the mean value increases with a constant speed $c$. This model was used before to better understand the statistics of athletic records (Ballerini and Resnick, \citeyear{Ballerini1985}; Gembris et al., \citeyear{Gembris2007}), but we showed that it is also capable of describing the occurrence of records in daily temperature recordings (Wergen and Krug, \citeyear{Wergen2010}; Rahmstorf and Coumou, \citeyear{Rahmstorf2011}). By considering (\ref{pngeneral}) one finds that for any constant drift $c$ the record rate in the LDM is of the following form:
\begin{eqnarray}\label{pncgeneral}
 P_n\left(c\right) = \int \mathrm{d}x\;f\left(x\right) \prod_{k=1}^\infty F\left(x+ck\right).
\end{eqnarray}
Most interesting for us is the statistics of records for a drift velocity $c$ much smaller than the width of the probability distribution $f\left(x\right)$. In most cases this width is just the standard deviation $\sigma$ of the probability distribution. Performing a series expansion of (\ref{pncgeneral}) in the regime of $c\ll n/\sigma$ one finds the following approximation for $P_n\left(c\right)$ \citep{Wergen2010}:
\begin{eqnarray}\label{pncapprox}
 P_n\left(c\right) \approx \frac{1}{n} + \frac{c}{\sigma}\frac{n\left(n-1\right)}{2}\int \mathrm{d}y\;[f\left(y\right)]^2 [F\left(y\right)]^{n-2}.
\end{eqnarray}
\cite{Franke2010} evaluated this expression for distributions from all three classes of extreme value statistics (Galambos et al., \citeyear{Galambos1994}; De Haan and Ferreira, \citeyear{DeHaan2006}; Nevzorov, \citeyear{Nevzorov2001}; Gumbel, \citeyear{Gumbel}). The dependence of the record rate $P_n\left(c\right)$ on the drift $c$ is systematically different between the three classes, but also within them one can find differences between different individual probability distributions. 
%A detailed study of $P_n\left(c\right)$ in the regime of $c\ll n/\sigma$ can be found in Franke et al. \citeyearpar{Franke2010}.

%For the Weibull class of probability distributions with a finite support the effect of the linear drift on $P_n\left(c\right)$ was found to be strong and increasing with $n$. For the distributions of the Fr\'echet class with power-law tails it decays with growing $n$ and vanishes for $n\rightarrow\infty$ \citep{Franke2010}. The behavior of $P_n\left(c\right)$ in the Gumbel class of distributions with an infinite support that decay faster than a power law is intermediate between these two cases. % For a simple exponential distribution with $f\left(x\right) = 1/\nu e^{-x/\nu}$ the effect of the drift on the record rate is 
%independent of $n$ and we have $P_n\left(c\right) \approx \frac{1}{n} + c\nu/2\sigma$. 

\begin{figure}
\includegraphics[width=0.48\textwidth]{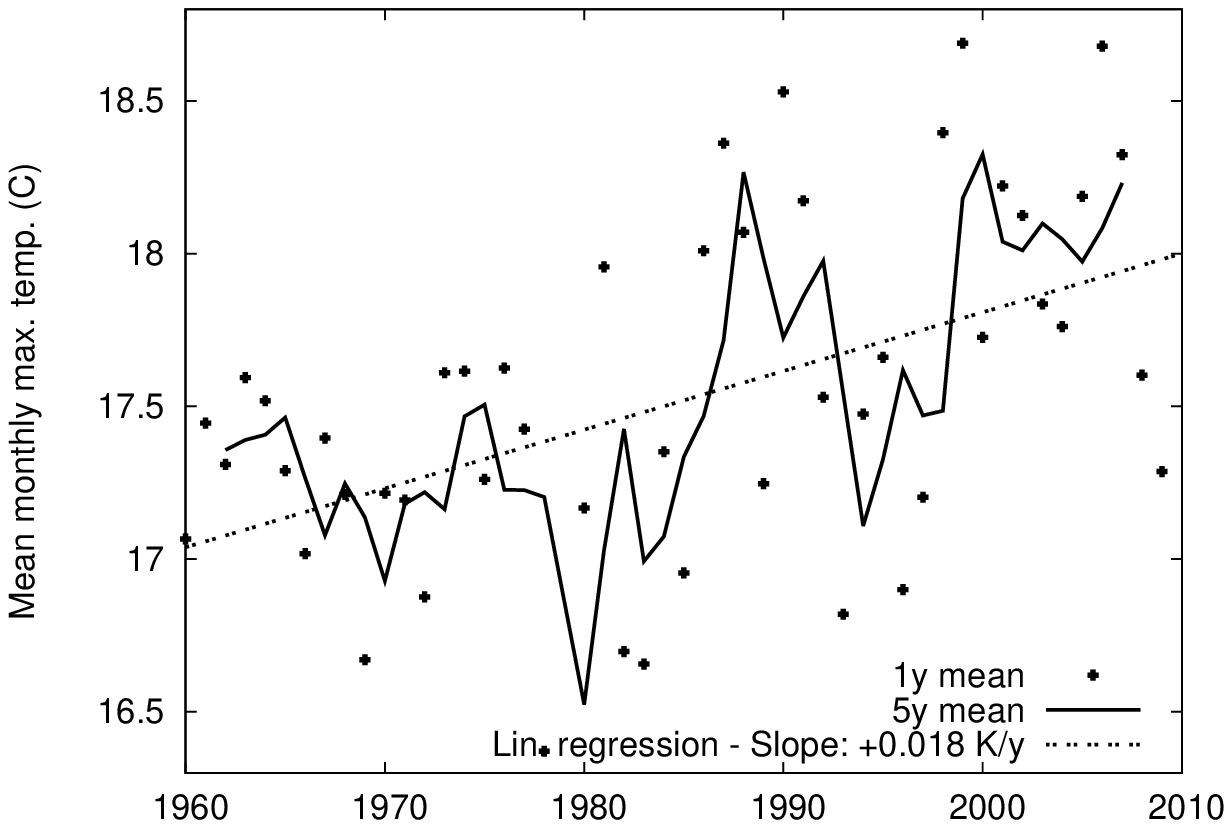}\\
\includegraphics[width=0.48\textwidth]{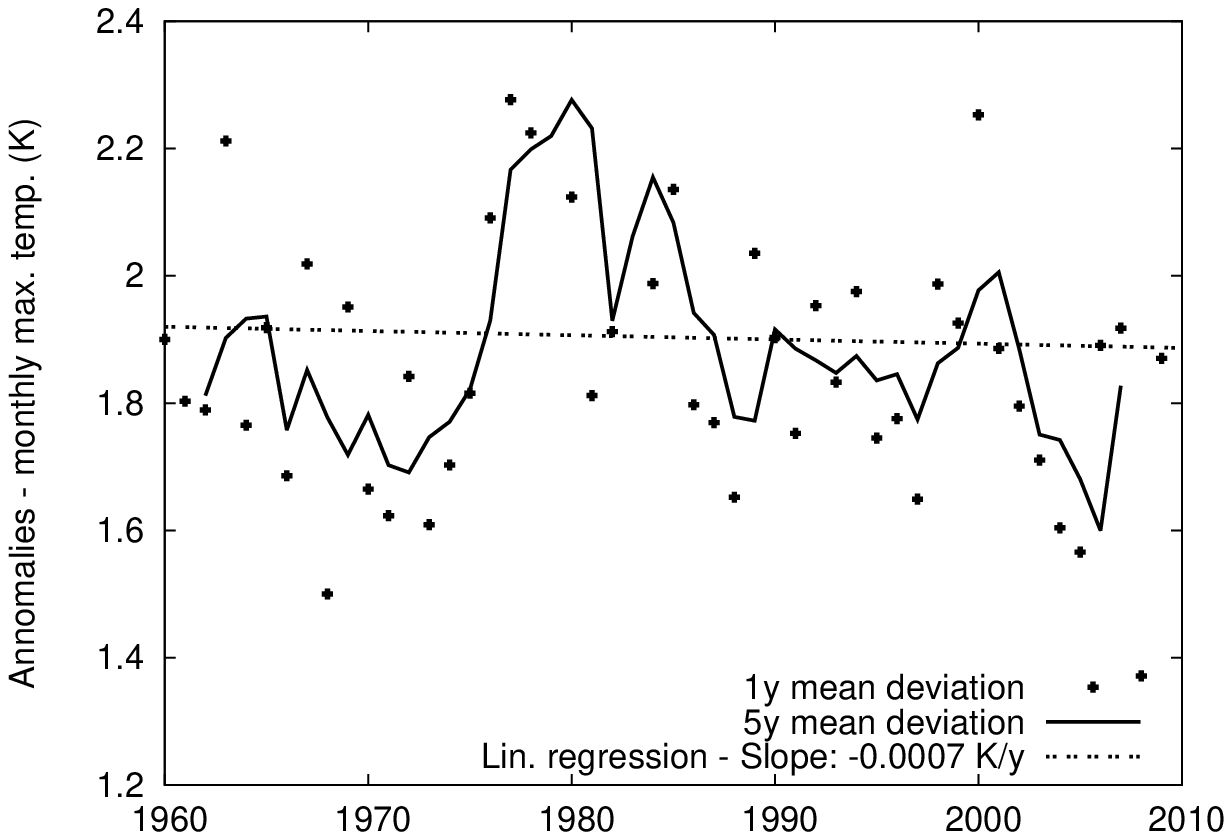}
\caption{\label{Fig:US_monthly_mean} Annual mean temperature (top) and
  standard deviation (bottom) of monthly mean temperatures from
  contiguous US stations 1960-2010. Dots are one year averages, full lines give a five year running mean. The anomalies defining the standard deviations of the mean temperature were computed by subtracting the mean and linear trend 1960-2010 individually for each month. Annual mean values increase over the last 50 years with an average rate of $0.018\;\textrm{K}^{\circ}/y$, whereas the standard deviation of the monthly anomalies remained more or less constant. }
\end{figure}

Most interesting for our applications is $P_n\left(c\right)$ for a Gaussian distribution of the following form:
\begin{eqnarray}\label{Gaussian_pdf}
 f\left(x\right) = \frac{1}{\sqrt{2\pi}\sigma}e^{-\frac{\left(x-\mu\right)^2}{2\sigma^2}}.
\end{eqnarray}
Here, $\mu$ is the mean value of the probability distribution and $\sigma$ its standard deviation. The approximate evaluation of Eq. (\ref{pncapprox}) for large enough $n$ with $c\ll n/\sigma$ and a Gaussian distribution with standard deviation $\sigma$ yields:
\begin{eqnarray}
 P_n\left(c\right) \approx \frac{1}{n} + \frac{c}{\sigma} \frac{2\sqrt{\pi}}{e^2}\sqrt{\textrm{ln}\left(\frac{n^2}{8\pi}\right)}.
\end{eqnarray}
\cite{Franke2010} compared this approximation to numerical simulations
and good agreement was found for $n>7$ and $c\ll n/\sigma$. The result was applied to the statistics of record-breaking temperatures for the first time by \cite{Wergen2010}, who showed that the regime of $c/\sigma\ll 1$ is appropriate for the modeling of daily temperatures in a climate with moderate warming like that occurring in the last decades. This result will also be employed for comparison with observational data in section 4 of this article. 

Another complication that can arise in the context of record
statistics of historical temperature recordings, is the problem of
rounding. The fact that, for technical reasons, temperature
measurements can only be recorded up to a certain degree of accuracy
opens up the possibility for ties. Usually the observations are
rounded down (or up) to a certain value $kd$, where $d$ is a
discretization scale (e.g. $0.1\;\textrm{K}^{\circ}$ if temperatures
are stored up to the first digit) and $k$ an integer
number. \cite{Wergen2012} recently explored the interesting and manifold consequences of this rounding for the statistics of records in time series of iid RV's in the context of the three universality classes of extreme value statistics.

\section{General introduction of the data}
%% alle Einheiten in MKS!!!
\subsection{US monthly station data}

We focus our analysis on two different sets of data from Europe and the contiguous United States. The purpose of this section is to analyze the distributional properties of these data sets. We want to know if the LDM discussed in the previous section is applicable to the observational data. In particular we will examine if a Normal or Gaussian probability density function with a slowly changing expectation value is a reasonable approximation. Therefore we are interested in the linear trend $c$ of the daily and monthly temperature averages and the standard deviation $\sigma$ around that transient mean value $\mu + ck$. As outlined above the most important quantity in our analysis is the normalized drift $c/\sigma$. 

\begin{figure}%\begin{table}
\includegraphics[width=0.48\textwidth]{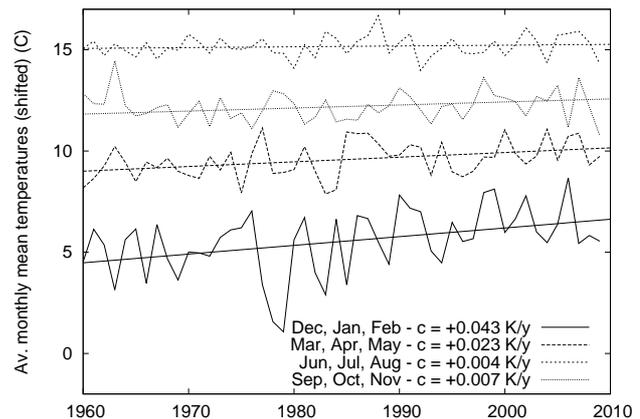}
\caption{\label{Fig:US_seasonal_mean} Time series of seasonal averages of monthly mean temperatures from contiguous US stations 1960-2010. For clarity, three year moving averages are plotted. Legend shows the estimated trends over the period 1960-2010. Winter months exhibit the strongest warming with $c=0.043\;\textrm{K}^{\circ}/y$, during spring and fall the warming is moderate with $c=0.023\;\textrm{K}^{\circ}/y$ and $c=0.007\;\textrm{K}^{\circ}/y$, and in the summer months we find only a very small warming of $c=0.004\;\textrm{K}^{\circ}/y$. However for the normalized drift and therefore the record-statistics, also the standard deviation is important, which is significantly smaller in the summer as well (see text).}

\begin{tabular}[t]{|l|c|c|}
  \hline
& $\overline{c}$ & $\overline{\sigma}$ \\
  \hline\hline
Winter & $0.043\pm0.015\;\textrm{K}^{\circ}/y$ & $2.45\pm0.012\;\textrm{K}^{\circ}$ \\
Spring & $0.023\pm0.008\;\textrm{K}^{\circ}/y$ & $1.68\pm0.006\;\textrm{K}^{\circ}$ \\
Summer & $0.004\pm0.007\;\textrm{K}^{\circ}/y$ & $1.70\pm0.006\;\textrm{K}^{\circ}$ \\
Fall & $0.007\pm0.005\;\textrm{K}^{\circ}/y$ & $1.56\pm0.005\;\textrm{K}^{\circ}$ \\
  \hline\hline
& $\overline{c/\sigma}$ & $\overline{c}/\overline{\sigma}$\\
  \hline\hline
Winter & $0.017\pm0.006\;y^{-1}$ & $0.017\pm0.006\;y^{-1}$ \\
Spring & $0.014\pm0.005\;y^{-1}$ & $0.013\pm0.005\;y^{-1}$ \\
Summer & $0.002\pm0.004\;y^{-1}$ & $0.003\pm0.004\;y^{-1}$ \\
Fall & $0.004\pm0.003\;y^{-1}$ & $0.004\pm0.003\;y^{-1}$\\
\hline
\end{tabular}
\caption{\label{Tab:US_seas_driftnorm} Averaged drift $\overline{c}$, averaged monthly standard deviation $\overline{\sigma}$ and averaged normalized drift $\overline{c/\sigma}$ for the contiguous US stations 1960-2010. While in winter and spring there is a significant and strong drift in the data, we find only little or no effects in summer and fall. The analysis of $\overline{c}/\overline{\sigma}$ agrees perfectly with the results for $\overline{c/\sigma}$.}
%\end{table}
\end{figure}

We obtained monthly mean temperatures from 1217 weather stations of the contiguous US \citep{USHCN}. The data cover the period 1895 and 2010, but but not all stations are complete. %Parts of the data were determined by re-analysis methods (see \citep{USHCN}). 
Therefore we decided to analyse the 50 year period 1960 to 2010 (data set USM) which is the period with least missing data. This way we could consider $1217\times365$ time series. Based on the discussion of Wergen and Krug \citeyearpar{Wergen2010}, we estimate that the number of effectively independent time series in this data set is much smaller than the total number of series. The number of independent stations is limited because of correlations both in space and in time. Following \cite{Wergen2010} we estimate not more than 20-25 independent stations and around 36 independent calendar days, leading to a total number of around 700-900 independent time series.

As one can see in Fig. \ref{Fig:US_monthly_mean} the monthly averages show a clear increase in the mean value with a drift of $c = 0.018\pm0.005\;\textrm{K}^{\circ}/y$. In contrast, the standard deviation of the monthly averages remains constant around $\sigma = 1.88 \pm 0.03\;\textrm{K}^{\circ}$, which is smaller than the standard deviation of the daily measurements of around $5\;\textrm{K}^{\circ}$ for the same period~\citep{Wergen2010}. Using a linear regression analysis we tried to detect a trend in the standard deviation, but only found a insignificant trend of $-0.0007\pm0.0021\;\textrm{K}^{\circ}$. Assuming a constant $\sigma$, we obtain a normalized drift $c/\sigma \approx 0.010\pm0.003\;y^{-1}$, which is almost as large as the $c/\sigma$ for the European daily data in \citep{Wergen2010}. 

The seasonal dependence of the normalized drift $c/\sigma$ is
presented in Fig. \ref{Fig:US_seasonal_mean}. During winter the US
mean temperature increased most strongly, while the trend for summer
is much smaller. However, important for the record statistics is the ratio $c/\sigma$. The normalized trends are listed in Tab. \ref{Tab:US_seas_driftnorm}. Based on the theoretical results we expect a strong effect of the slow temperature increases on the record rates in winter and spring in contrast to summer and fall if all prerequisites of the theory are fulfilled. This is due to the strong difference in normalized drift 
(Tab. \ref{Tab:US_seas_driftnorm}, third column).

\subsection{European daily data}

\begin{figure}
\includegraphics[width=0.48\textwidth]{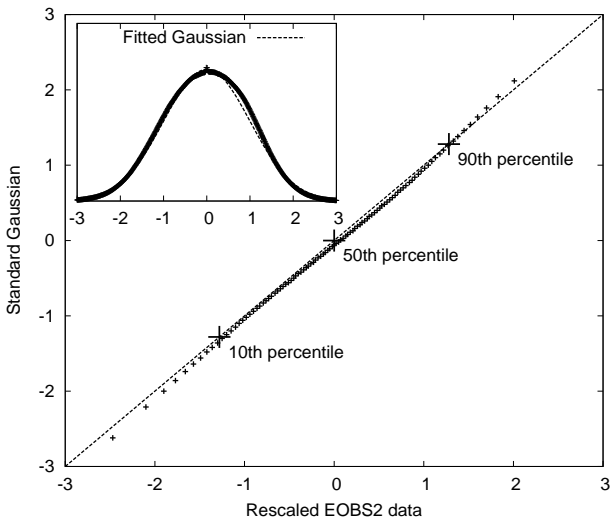}
\caption{\label{Fig:EUR_st_daily_distr} Q-Q plot of the linearly detrended and normalized daily maximum temperatures in the EOBS data (EOBS2) with RV's sampled from a Gaussian distribution with standard deviation unity. The $k$th percentile is the smallest value, which is larger than the smallest $k$ per cent of all measurements. The percentiles collapse on a line showing that the rescaled measurements are Gaussian to a good approximation. The inset shows the estimated probability density of the data compared to a Gaussian (dashed line). }
\end{figure}

The second data set we obtained is taken from the E-OBS  project~\citep{EOBS}. The E-OBS data set provides daily minimum, mean and maximum temperatures on a 0.25 degree regular grid for most of Europe and also parts of northern Africa starting from 1950 to the end of 2010. We analyzed the daily maximum temperature data set for two time periods, one between 1950 and 2010 (data set EOBS1) and a second one between 1980 and 2010 (data set EOBS2). Again, it is important to mention, that despite the fact that we analyzed a much larger number of time series, only around 300-400 series were effectively independent. This number is smaller in Europe due to the smaller size of the considered area. The time series analysis of the mean of the daily maximum temperature and the standard deviation were not significantly different from the station data considered by Wergen and Krug \citeyearpar{Wergen2010}. During the first 30 years of EOBS1 the mean temperature shows no prominent trend. During the second period, between 1980 and 2010 (EOBS2), 
we find a clear warming trend of about $c=0.042\;\textrm{K}^{\circ}/y$. In data set EOBS2, the standard deviation of the daily temperatures  around the linear trend remained constant at a value of about $3.4\;\textrm{K}^{\circ}$ resulting in a normalized drift of around $c/\sigma\approx0.012\pm0.003\;y^{-1}$, which is slightly smaller than that obtained from European station data \citep{Wergen2010}. 

In Fig. \ref{Fig:EUR_st_daily_distr} we present a quantile-quantile (Q-Q) analysis of the distribution of the detrended and normalized temperature measurements in data set EOBS2. We subtracted a linear trend, which was obtained by a regression analysis, from the time-series at each grid point and then normalized them by dividing with the standard deviation of the measurements in the individual series. Then the percentiles of the measurements were plotted against the corresponding percentiles of a Gaussian distribution with mean zero and standard deviation unity. The first percentile is defined as the value which is larger than the smallest one per cent of all data points. Correspondigly, the $k$th percentile is the smallest value, which is larger than the smallest $k$ per cent of all measurements. Since the percentiles of the observational data and of the Gaussian distribution are almost identical, the quantile-quantile plot shows that the measurements are Gaussian to good approximation. The data therefore 
fits our model of a Gaussian with a linear drift. The inset shows an estimation of the probability density function based on a simple histogram estimator with a bin width of $0.01 \;\textrm{K}^{\circ}$.

\section{Occurrence of records in daily and monthly temperatures}

\subsection{US monthly station data}

\begin{figure}
\includegraphics[width=0.48\textwidth]{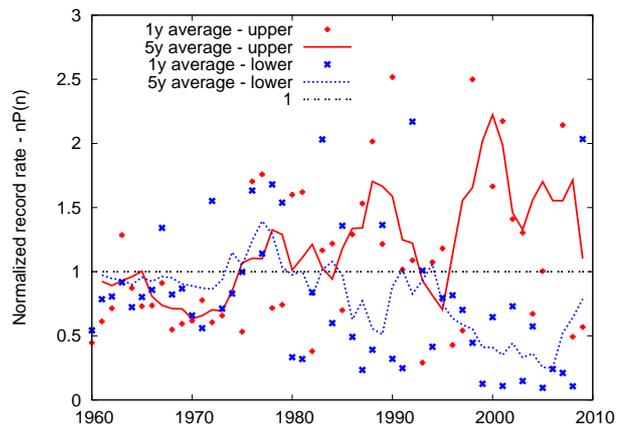}
\caption{\label{Fig:US_rec_rate_5009} Normalized record rate $nP_n$ of monthly mean temperature data from US stations. The dots represent the annual averages of $nP_n$ and the lines represent a five-year running average. The line at one is the behaviour expected in the case of a stationary climate with realizations from an iid process. Before 1980 it is hard to distinguish between the upper and lower record rate. After 1980 the upper record rate increases and the lower record rate decreases. }

\begin{tabular}[t]{|l|c|c|}
  \hline
& Rec. ratio - US stations & Ratio - LDM\\
  \hline\hline
Winter & $7.6$ & $7.1\pm5.6$ \\
Spring & $9.4$ & $4.8\pm3.7$ \\
Summer & $2.4$ & $1.2\pm0.7$ \\
Fall & $1.6$ & $1.5\pm0.4$ \\
  \hline
\end{tabular}
\caption{\label{Tab:US_rec_ratios} Ratio between the number of upper and lower records after $n=50$ years in the monthly data from the US stations along with the corresponding predictions from the LDM. The values for the normalized drift $\overline{c/\sigma}$ were taken from Tab. \ref{Tab:US_seas_driftnorm}. The upper and lower record numbers were averaged over the last $10$ years of the observation period. }

\end{figure}

\cite{Wergen2010} showed that the record rate $P_n$ of daily US
station temperatures does not display a significant effect of the
slowly evolving mean temperature. When one considers monthly data this changes basically due to the reduced variance. In Fig. \ref{Fig:US_rec_rate_5009} we show the normalized record rate $nP_n$ for monthly means. Apparently, for the years after 1980, the number of high temperature (upper) records is above the stationary case of $nP_n=1$ and the number of low temperature (lower) records is decreased. Compared to the null-hypothesis of a stationary climate, in the years between 2000 and 2010, we found around 1.52 times the number of hot temperature records and only half (0.48) the cold records one would expected. Therefore, the number of upper records was about 3.13 times the number of lower records. 

These results are in good agreement with the Gaussian LDM we described above. For the normalized drift $c/\sigma \approx 0.01 \;y^{-1}$ obtained from our data, the LDM predicts $1.44\pm0.13$ times more upper records than expected in a stationary model and only $0.50\pm0.15$ the number of lower records. The ratio between upper and lower records in the last 10 entries should be $2.88\pm1.12$, which is also in agreement with the observational data. Interestingly, without the very cold year of 2009, these numbers become more extreme. Ignoring the 2009 data, the ratio between the number of upper and lower records in the last ten years is around 4.9.

Additionally we can consider the record rate for the different
seasons. We computed the ratios between upper and lower records for
the four seasons and compared them to the estimates of $c/\sigma$ and
the resulting analytical predictions from the LDM with Gaussian
RV's. The results of this analysis are shown in
Tab. \ref{Tab:US_rec_ratios}. As expected mostly the winter and spring months experience a strong effect of slow temperature increase on the record statistics. In spring a heat record during the last ten years of the observational period was almost ten times as likely as a cold record. In winter this factor is almost eight. However, also in summer and fall the moderate warming lead to a significant effect in the statistics of records. Interestingly in spring and summer the ratio between the upper and lower record rate is larger than predicted by the Gaussian - LDM, but given the large fluctuations in the data, this can very well be a coincidence. It is also interesting to notice, that both in winter and in spring less then 20 \% of the cold records that we would have expected in the case of a stationary climate actually occurred.

\begin{figure}
\includegraphics[width=0.45\textwidth]{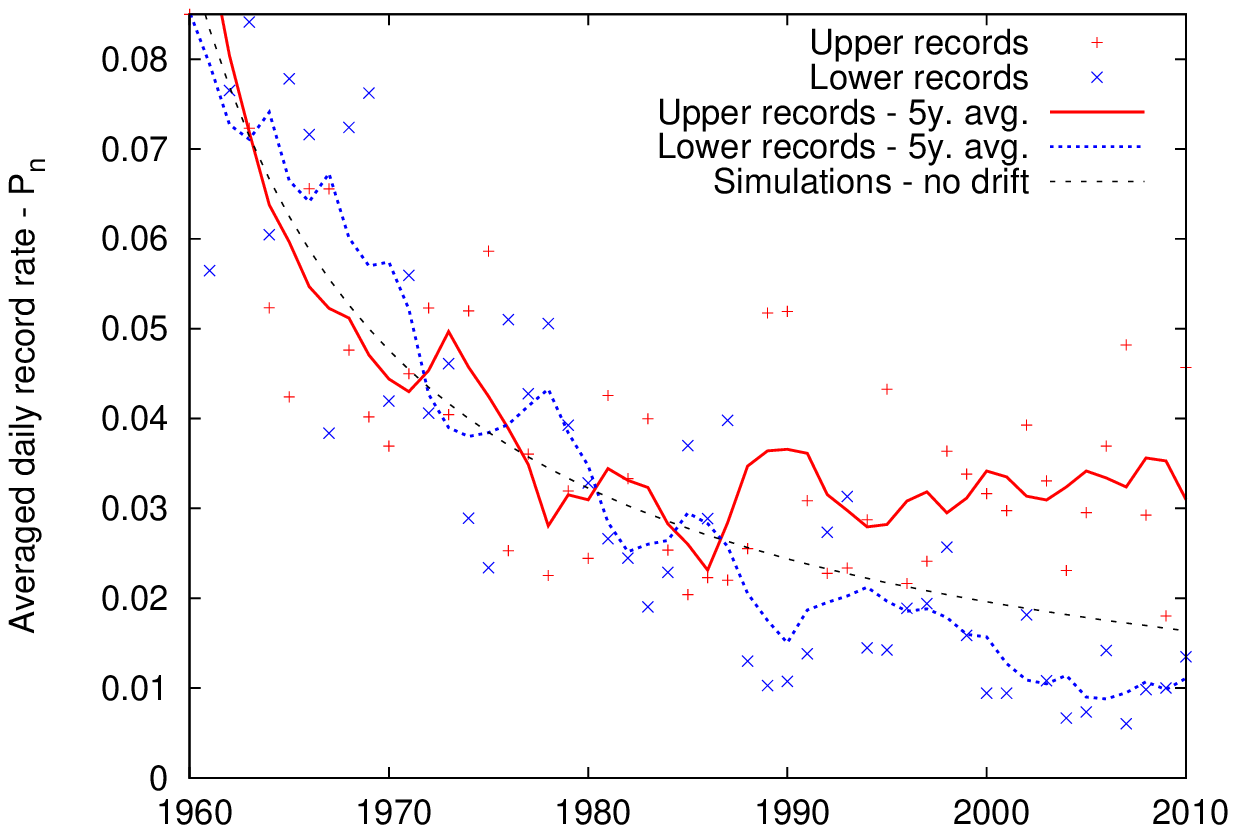}\\
\includegraphics[width=0.45\textwidth]{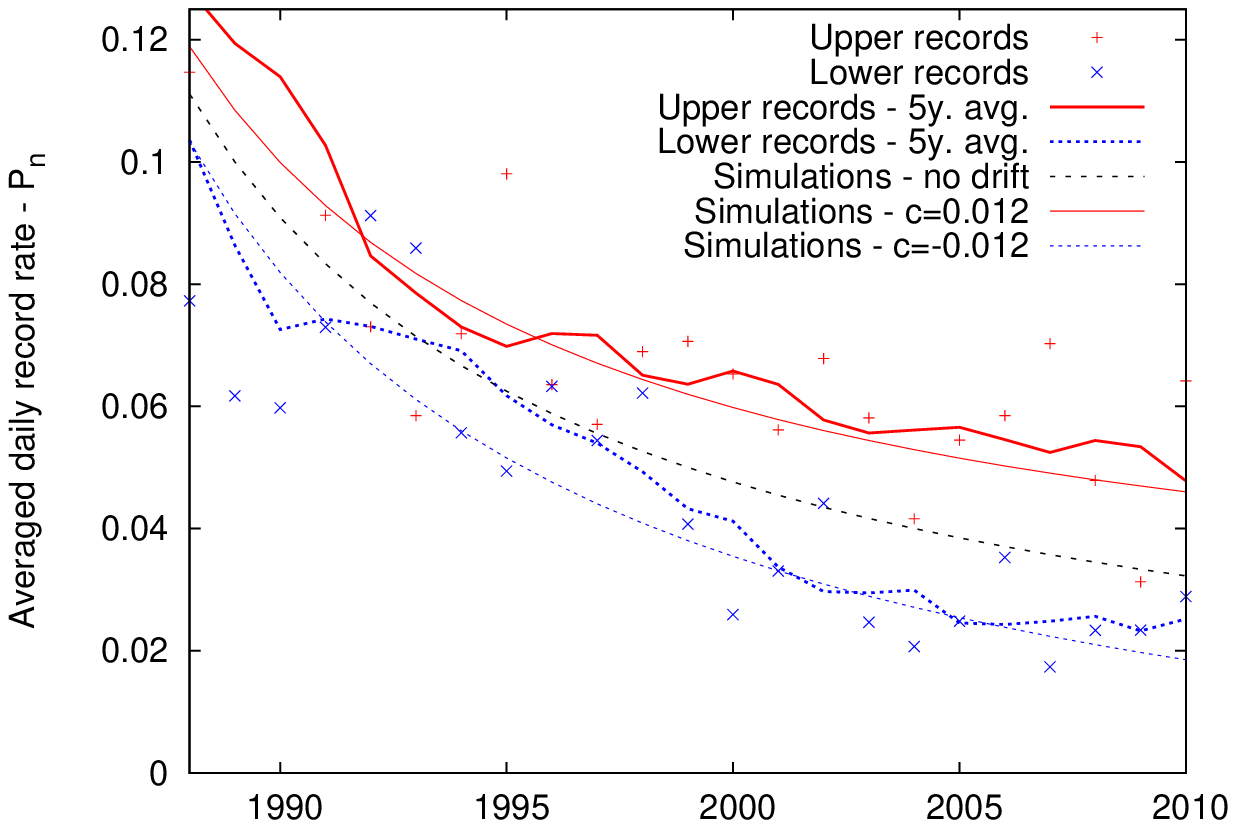}
\caption{\label{Fig:EOBS_Pn} \textbf{Upper figure:} Record rate per station and calendar day in the daily data set EOBS1 (1950-2010). Red represents the occurrence of upper records and blue the occurrence of lower records. The crosses give the annual average and the lines a five-year running average. The black dashed line is the $1/n$ behavior we would have expected in the case of a stationary climate. Until 1980 the upper and lower record rate are hard to distinguish and do not vary much from the stationary behavior. After 1980 the rate of upper records ceases to decrease and approaches a constant value of about 0.033 (12 records per year), while the rate of lower records decreases faster than the $1/n$ line. \textbf{Lower figure:} The same analysis for EOBS2 (1980-2010). Here the upper record rate is significantly increased compared to the no-warming model and the lower record rate. Red and blue dashed lines are the analytical predictions with a normalized annual mean trend of $c/\sigma = 0.012 y^{-1}$ estimated from the 
data.}
\end{figure}

\subsection{European daily re-analysis data}

We analyzed the record rates for the two time spans chosen for EOBS1 and EOBS2 (1950-2010 and 1980-2010) and compared them to the analytical predictions from our linear drift model. The results can be found in Fig. \ref{Fig:EOBS_Pn}. Considering the data set EOBS1 we find that in the first 30 years of the observation period, there was no significant effect of slowly increasing temperatures on the record statistics. Between 1950 and 1980, both the rate of upper and lower records behaved roughly like the record rate of iid RV's. After 1985 the upper record rate is significantly larger than the lower record rate, with the lower record rate significantly decreasing. For the years after 2000 the ratio between upper and lower records exceeds two and even approaches a value of three at the end of the observational period. 

In the data set EOBS2, the averaged upper record rate was higher than the lower record rate for almost the entire time span between 1980 and 2010. At the end of the observational period there were twice as many upper records as there were lower records. Here, the predictions from our linear drift model are very accurate in predicting the effect of the warming on the occurrence of both upper and lower records.

\begin{figure}
\includegraphics[width=0.48\textwidth]{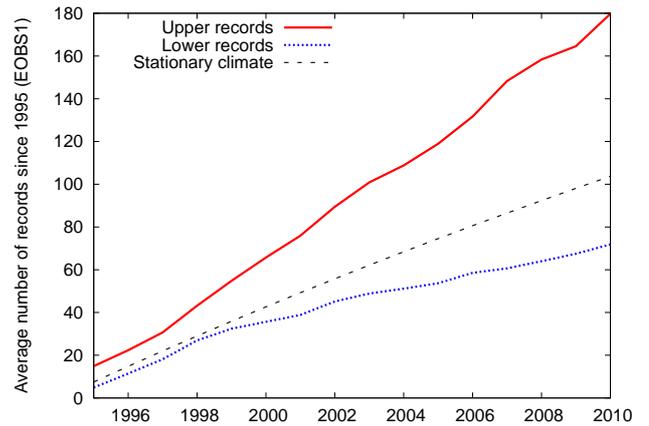}
\caption{\label{Fig:Records_EOBS1_since_1995} Excess number of records that occurred since the year 1995 per grid point per year in data set EOBS1. The record temperatures were recorded from the beginning of the time series. For this figure we only summed up records that occurred after the beginning of 1995. The records that contribute to this figure are the ones that exceeded the record that was valid at the end of 1994.}
\end{figure}

In Fig. \ref{Fig:Records_EOBS1_since_1995} we considered the number of records that occurred over a prolonged time span. We analyzed the upper and lower record temperatures that occurred in data set EOBS1 and started summing up the record rate beginning with the year 1995. Even though this figure is difficult to compare with our LDM, since we do not assume a linear drift over the entire time span from 1950 to 2010, it shows how strong the effect of warming on the record occurrence was in the last decades. Towards the end of the considered period, one finds that upper records occurred on average more than $2.5$ times more often than lower records.

\section{Theory: Distributions of record values}

After discussing the occurrence of record events in the previous sections, we now turn to the record values themselves. There are in principle two ways to study the effect of a slowly changing mean value on the record values. One approach is to consider the probability density function (pdf) of record values with a fixed record number $k$, which can however happen at an arbitrary time $n$. The other is to study pdf's of record events occurring at a fixed time step $n$. We do not know any simple way of computing the pdf's of the $k$th records in the presence of linear drift analytically. We will discuss the iid case and present results of some numerical simulations. If we consider instead the probability densities of record values for records occurring at a fixed time $n$ we can use the methods described above to compute a small $c$ approximation in the framework of the LDM.

\subsection{Records in the n'th step}

With the general expression for the record rate $P_n\left(c\right)$ in the LDM we gave in section 2, we can also obtain an expression for the probability distribution or cumulative distribution (cdf) 
\begin{eqnarray}
Q_n\left(c,x\right) = \textrm{Prob}\left[X_n\textrm{ is rec. \& } X_n < x\right]
\end{eqnarray}
 of a record that occurs in a certain time step $n$. This is given by:
\begin{eqnarray}
 Q_n\left(c,x\right) = \frac{1}{P_n\left(c\right)} \int^x \mathrm{d}y\;f_X\left(y\right) \prod_{k=0}^{n-1} F_X\left(y+kc\right).
\end{eqnarray}
The prefactor $P_n\left(c\right)^{-1}$ is necessary for the normalization. Then, the corresponding pdf of a record that occurs at time $n$ is given by $p_n\left(c,x\right) = \frac{\mathrm{d}}{\mathrm{d}x} Q_n\left(c,x\right)$.
%\begin{eqnarray}
% p_n\left(c,x\right) = \frac{1}{P_n\left(c\right)} \;f_X\left(x\right) \prod_{k=0}^{n-1} F_X\left(x+kc\right).
%\end{eqnarray}
For the iid case ($c=0$) this reduces to the well known pdf of the maximum of $n$ iid RV's: $p_n\left(0,x\right) = n f_X\left(x\right) F_X^{n-1}\left(x\right)$.
%$Q_n\left(0,c\right) = n\int^x \mathrm{d}y\; f_X\left(y\right) F_X^n\left(y\right)$ and therefore a pdf 
Most interesting and important for our analysis of temperature record values is the mean value $\mu_n\left(c\right)$ of a record that occurs at time $n$. This is given by the first moment of $p_n\left(c,x\right)$: $\mu_n\left(c\right) = \int \mathrm{d}y\;y\;p_n\left(c,x\right)$.
%\begin{eqnarray}
% \mu_n\left(c\right) = \frac{1}{P_n\left(c\right)} \int \mathrm{d}y\;y f_X\left(y\right) \prod_{k=0}^{n-1} F_X\left(y+kc\right).
%\end{eqnarray}
Similar to the case of the record rate $P_n\left(c\right)$ we can compute a series expansion for this expression in the regime of $cn\ll\sigma$. Doing this we find
\begin{eqnarray}
 \mu_n\left(c\right) \approx \left(1-\frac{c}{2}n^3 I_n^{\left(0\right)}\right)\mu_n\left(0\right) + \frac{c}{2}n^3 I_n^{\left(1\right)},
\end{eqnarray}
where we defined $I_n^{\left(j\right)} := \int \mathrm{d}y\;y^j [f_X \left(y\right)]^2 [F_X\left(y\right)]^{n-2}$ and
$\mu_n\left(0\right) = n \int \mathrm{d}y\;y f_X\left(y\right) [F_X\left(y\right)]^{n-1}$.
Note that $I_n^{\left(0\right)}$ is the integral that appears in our result for the record rate $P_n\left(c\right)$ [Eq. (\ref{pncapprox})]. Furthermore $\mu_n\left(0\right)$ is the mean value of a record that occurs at time $n$ in the case of iid RV's. %Since we have an exact expression for the pdf $p_n\left(c,x\right)$, we can of course also compute higher moments and, for instance, the variance of a record that occurs in the LDM at a certain time $n$. For the variance we find:
%\begin{eqnarray}
% \sigma_n^2\left(c\right) & \approx & \sigma_n^2\left(0\right) \left(1-\frac{c}{2}n^3 I_n^{\left(0\right)}\right) \nonumber \\ & & + \frac{c}{2} n^3 \left(\mu_n\left(0\right)\left(I_n^{\left(0\right)}-I_n^{\left(0\right)}\right) + I_n^{\left(2\right)}\right)
%\end{eqnarray}
\begin{figure}
\includegraphics[width=0.48\textwidth]{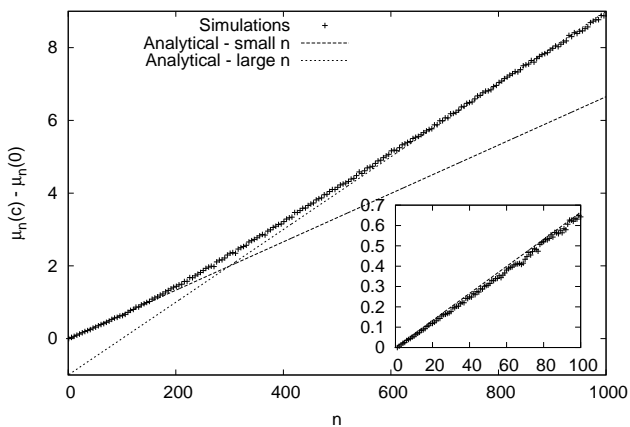}
\caption{\label{Fig:Mean_LDM} Effect of the drift on the mean value $\mu_n\left(c\right)- \mu_n\left(0\right)$ for Gaussian RV's. The crosses are results from numerical simulations. For $c=0$ and $c=0.01$ we performed $10^6$ runs each with $n=1000$ RV's and averaged the differences between the mean record values. The lines are the analytical results for small and large $n$. The inset shows a comparison with the small $n$ result for $n$ between $0$ and $100$. 
%Both in the small and in the large $n$ regime the analytical predictions describe the behavior of $\mu_n\left(c\right)- \mu_n\left(0\right)$ very accurately.
}
\end{figure}
%From now on we will focus on the mean value $\mu_n\left(c\right)$. 
Again, as in the work of \cite{Franke2010} it is possible to compute $\mu_n\left(c\right)$ for instances of all three classes of extreme value theory and we find a systematic classification of the behavior with respect to these classes~\citep{WergenUnpup}. Here, we will focus on the Gaussian density, because it is the one we need for our comparison of the LDM to observational data. For the same Gaussian pdf as in section 2, we find that:
\begin{eqnarray}
 \mu_c\left(c\right) \approx \mu_n\left(0\right) + n \frac{c}{\sigma}\frac{2\sqrt{\pi}}{e^2}\textrm{ln}\left(4\right).
\end{eqnarray}
Interestingly, in contrast to the case of the record rate $P_n\left(c\right)$, the effect of the drift on $\mu_n\left(c\right)$ up to first order in $c$ is linear in $n$. In the limit of $cn\gg\sigma$ we expect that $\mu_n\left(c\right) - \mu_n\left(0\right)$ is again linear in $n$. It is easy to see that for $n\rightarrow\infty$ we get $\mu_n\left(c\right) - \mu_n\left(0\right) \approx cn$. In this regime, the drift dominates the behaviour and the mean record value $\mu_n\left(c\right)$ is given by the linearly growing mean of $f\left(x-cn\right)$ plus a sublinear contribution.

We compared these analytical findings for the Gaussian density to numerical simulations. The results can be found in Fig. \ref{Fig:Mean_LDM}. Both in the small and in the large $n$ regime the analytical predictions describe the behaviour of $\mu_n\left(c\right)- \mu_n\left(0\right)$ very accurately. For the chosen drift rate of $c=0.01$ the intermediate regime of $cn\propto\sigma$, where both descriptions given above fail seems to be very small.

\subsection{k'th records}

\begin{figure}
\includegraphics[width=0.48\textwidth]{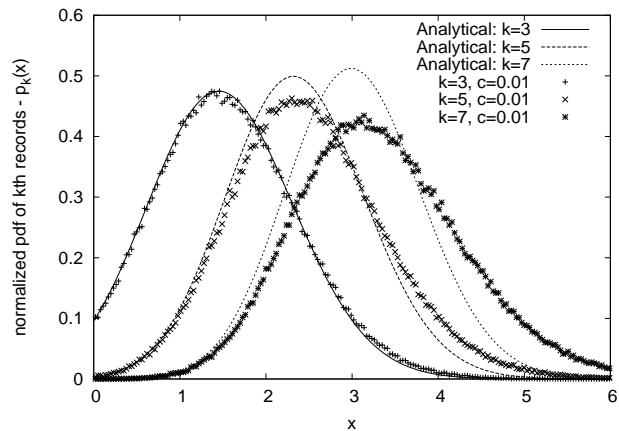}
\caption{\label{Fig:k_value_distr_LDM} Normalized distributions of
  records with record number $k$ occurring at an arbitrary time $n$
  for Gaussian RV's in the iid case and with a constant drift of
  $c=0.01$. Shown are the pdf's for $k=3,5$ and $7$. The analytical
  results for the iid case (Eq. (\ref{p_k})) are plotted as lines, the
  crosses are the numerical results with drift. For each $k$ and $c$
  we averaged over $10^6$ realizations of a time series of length
  $10^6$. The finite length of the time series does not have a significant effect on the distributions of records with the given values of $k\ll \ln 10^6 + \gamma$.}
\end{figure}

In the case of the record values of a record that occur at an arbitrary time $n$, but with a fixed record number $k$, one can easily give the full distribution of record values in the iid case. Here, we will briefly discuss the findings presented by Arnold et al. \citeyearpar{Arnold1998}. There it was shown that the pdf $p_k\left(x\right)$ of a record that occurs with record number $k$ can be written as follows:
\begin{eqnarray}\label{p_k}
 p_k\left(x\right) = \frac{f_X\left(x\right)}{\left(k-1\right)!}\left(-\textrm{ln}\left(1-F_X\left(x\right)\right)\right)^{k-1}.
\end{eqnarray}
This result is basically a consequence of the lack-of-memory property of the exponential distribution, i.e. the fact that a new record from an exponential distribution, independent from the time of its occurrence, will always be an exponential RV plus the value of the last record \citep{Arnold1998}. This leads to the results that for an exponential distribution the $k$th record has a Gamma distribution $p_k^{exp}\left(x\right) = \Gamma\left[k-1,x\right]$. The general result given above can then be obtain through a simple mapping.

Unfortunately, for RV's with a time-dependence like a linear drift, the lack-of-memory property is lost and we do not know how to obtain a simple expression for the $p_k\left(x\right)$ for the LDM. In Fig. \ref{Fig:k_value_distr_LDM} we show the normalized densities of $k$th records for a few small record numbers $k$. Apparently, already for a small drift of $c=0.01$, there is a significant effect on the record value pdf for larger $k$. Also the width of the pdf increases as an effect of the drift. It seems that especially the right tails of the distributions become broader. In the context of temperature records these events in the tails are particularly interesting. The simulations in Fig. \ref{Fig:k_value_distr_LDM} agree qualitatively with the results presented in Fig. 5 in the work of Redner and Peterson \citeyearpar{Redner2006}, although in this publication the effect of the drift on the distributions is weaker due to a smaller ratio of $\frac{c}{\sigma}$. 

\section{Distributions of record values in European temperature recordings}

\begin{figure}
\centerline{\includegraphics[width=0.48\textwidth]{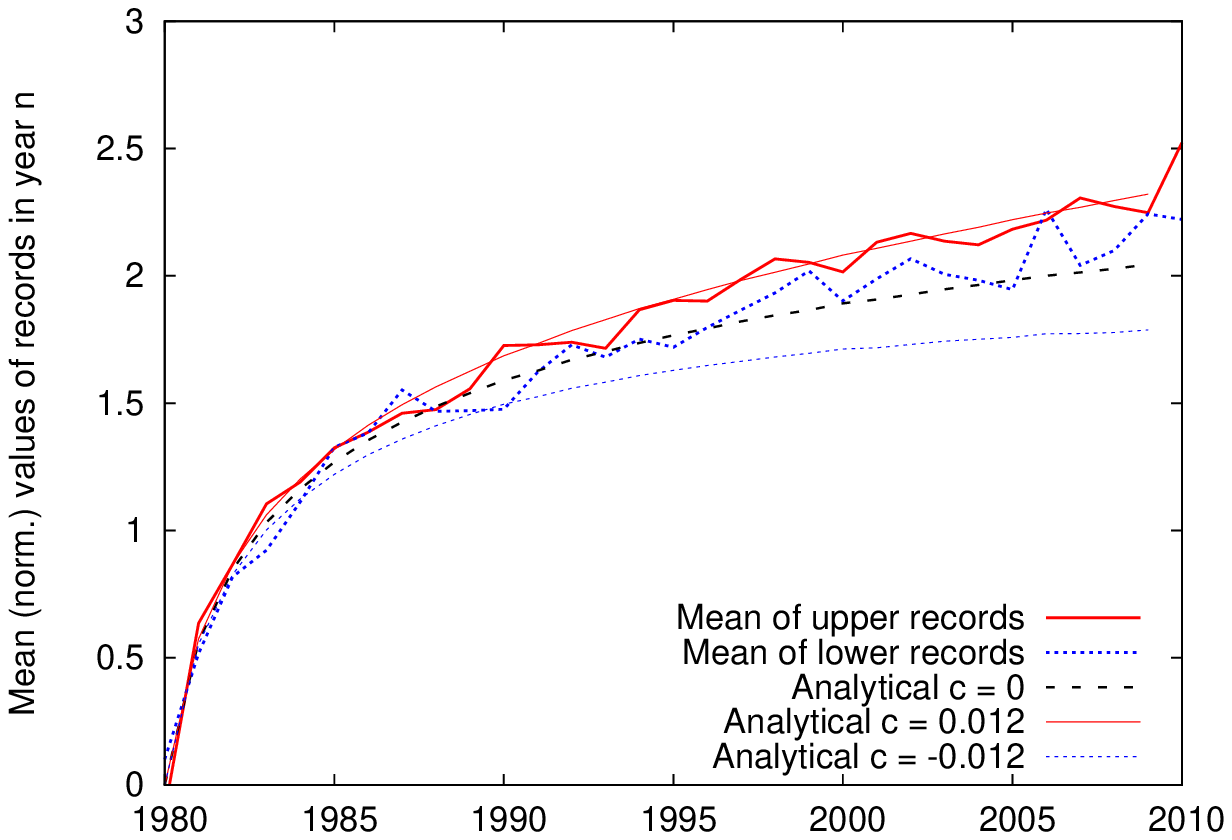}}
\centerline{\includegraphics[width=0.48\textwidth]{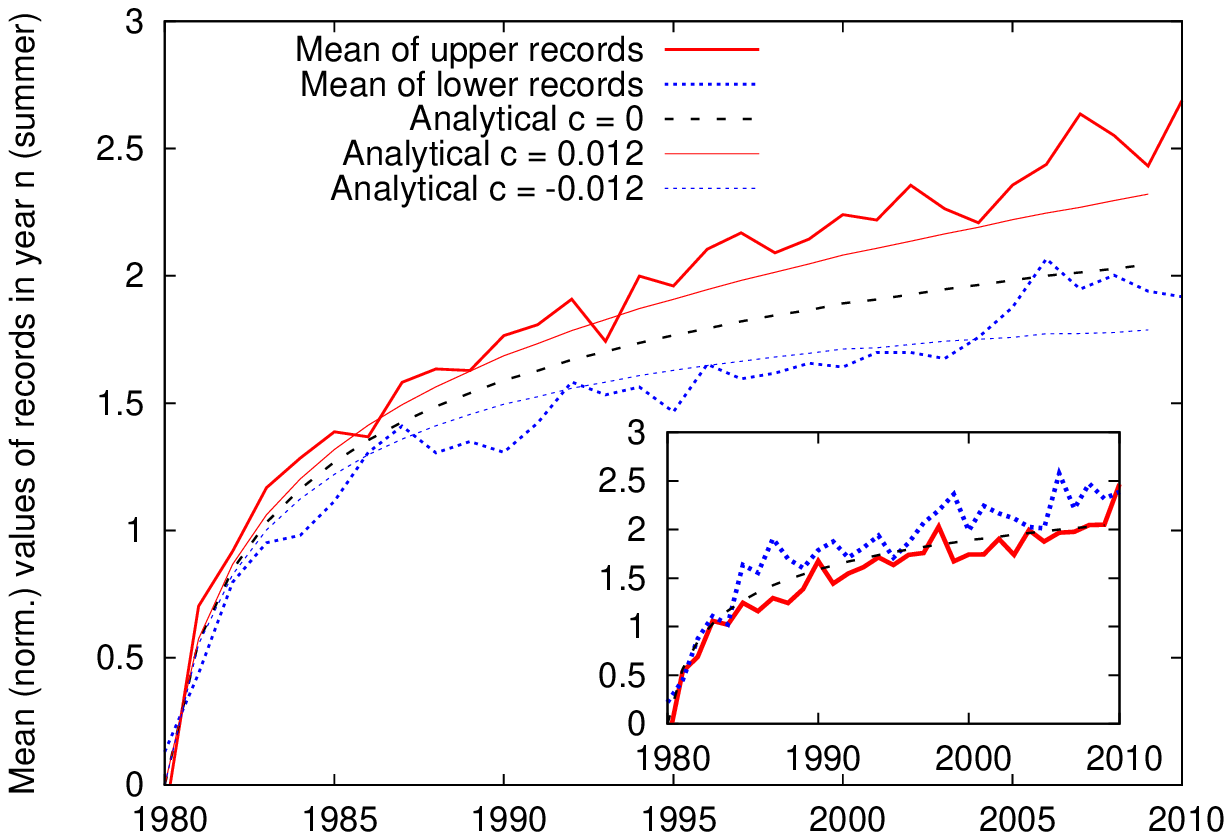}}
\caption{\label{Fig:EOBS_mean_rec_at_n} \textbf{Upper figure:} Time series of the rescaled mean value of temperature records that occurred in data set EOBS2. The full line gives the behaviour of the upper record value and the dotted line gives the inverse (negative) behaviour of the lower record. The values of the lower records were multiplied with $-1$ to make them comparable to the upper records. The rescaling is described in more detail in the text. The figure also contains the analytic results for no drift ($\tilde{c}=0$) and the LDM with a normalized drift of $\tilde{c}=0.012\;y^{-1}$ that was estimated from the observations. \textbf{Lower figure:} The same plot, but only for the two summer months of July and August. The analytic results for $\tilde{c}=0$ and the LDM case $\tilde{c}=0.012\;y^{-1}$ are added. The inset shows the same analysis for the two winter months January and February. }
\end{figure}

Based on the analytical work from the previous section we can now consider the values of records in observational data. For this analysis we focused on the European reanalysis data and in particular EOBS2. 

Here, the situation is a bit more complicated than in the case of the record rate $P_n\left(c\right)$. While the record rate in the LDM only depends on the normalized drift $c/\sigma$, the values of the records depend on the standard deviation $\sigma$ itself. The standard deviations of daily temperatures vary both spatially and seasonally and therefore it is difficult to compare the values of record breaking temperatures without some additional assumptions. To make the time series in EOBS2 more comparable we performed a rescaling of the data. This was done as follows:

The LDM assumes that an individual series of temperatures measurements $T_1,...,T_n$ measured in $n$ subsequent years is given by
\begin{eqnarray}
 T_k = \mu_0 + ck + \sigma \xi_k,
\end{eqnarray}
where the $\xi_k$ are iid RV's from a Gaussian distribution with standard deviation one. We subtract the intercept $\mu_0$ and divide by the standard deviation $\sigma$ to obtain the following time series
\begin{eqnarray}\label{rescaling}
 \tilde{T}_k = \frac{1}{\sigma}\left(T_k - \mu_0\right).
\end{eqnarray}
It is easy to see that these rescaled measurements have standard deviation unity around a normalized linear drift $\tilde{c}\equiv c/\sigma$. This kind of rescaling was done for all time series in the respective data sets individually so that we obtained comparable series of rescaled measurements. This way the data is most-suitable for comparison with a Gaussian LDM. If the observations were perfectly uncorrelated Gaussian RV's with an arbitrary but fixed standard deviation the record values of the time series after this rescaling would look exactly like the record values from a Gaussian LDM with standard deviation one. These rescaled temperatures should then obey the following LDM:
\begin{eqnarray}
 \tilde{T}_k = \frac{c}{\sigma} k + \xi_k = \tilde{c}k + \xi_k.
\end{eqnarray}
It is important to notice that the ordering of the measurements and in particular the statistics of records is not altered by this procedure because
\begin{eqnarray}
 T_n = \max\left[T_1,...,T_n\right] \Rightarrow \tilde{T}_n = \max\left[\tilde{T}_1,...,\tilde{T}_n\right].
\end{eqnarray}
So if and only if $T_n$ was a record in the original series, $\tilde{T}_n$ will also be a record in the rescaled series. The record rate $P_n$ and the record number $R_n$ are therefore completely invariant under this rescaling and only the values of the records will change according to Eq. (\ref{rescaling}). 

\subsection{Mean values of records in a given year}

\begin{figure}
\centerline{\includegraphics[width=0.48\textwidth]{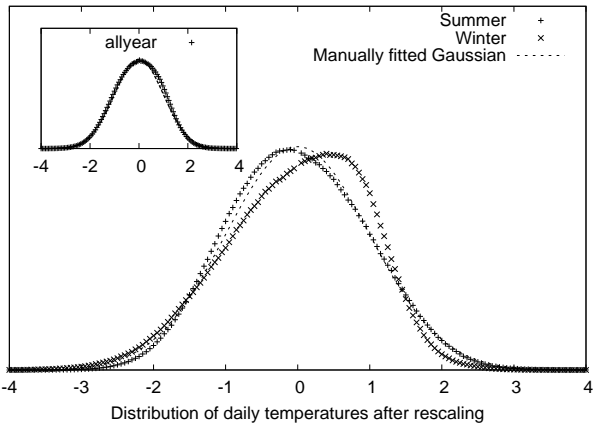}}
\caption{\label{Fig:Norm_distr_winter_summer} Estimated probability density functions of the daily temperature measurements in data set EOBS2 after rescaling to standard deviation unity. Estimation is based on a histogram. The main figure gives the pdf's for the two summer months July and August, as well as the two winter months January and February. The dashed line is a standard normal distribution. The pdf for the summer months is in good agreement with the Gaussian. The pdf for winter deviates significantly from a Gaussian. The asymmetry in winter is particularly interesting and explains the interesting behavior of the mean record values in the winter months (see main text). The inset shows the same analysis for the entire calendar year. Here, the rescaled distribution of daily temperature is again in good agreement with the Gaussian.}
\end{figure}

In Fig. \ref{Fig:EOBS_mean_rec_at_n} we analyze the mean values of a record that occurred in a certain year for the 30 years of observation in data set EOBS2. The rescaled data was first analyzed for each time series individually and then averaged over all grid points and calendar days. The upper figure gives the behavior of the mean value for the entire calendar year. The figure also shows the analytic results for the iid case and for the LDM with a normalized drift of $\tilde{c}=0.012$, which is determined from the observations. The behaviour of the mean upper and lower record values appear to be very similar. Both curves seem to have exactly the same shape and both lie slightly above the null hypothesis of a stationary climate with Gaussian daily temperatures. These results are not in agreement with the analytic results given by the LDM and do not show any apparent effect of slow temperature increase on the statistics of record values. 

The lower half of Fig. \ref{Fig:EOBS_mean_rec_at_n} shows the same analysis but only for two months in summer (July and August) as well as for two months in winter (January and February) in the inset. In summer, we find a much better agreement of the observations with a Gaussian LDM. The mean of the upper records increases much faster than the negative mean of the lower records, i.e., the upper records are more extreme than the lower records. Here, the difference between upper and lower records is in good agreement with the difference predicted from the LDM. In the two winter months (inset), the situation is completely different. Here, the negative mean of the lower records increases faster than the mean of the upper records. It seems, that despite a significant positive trend in the mean values of the daily temperatures, the values of the lower records in winter are still  more extreme than those related to the upper records. %The values of lower records are further away from the linearly increasing mean of 

\begin{figure}
\centerline{\includegraphics[width=0.48\textwidth]{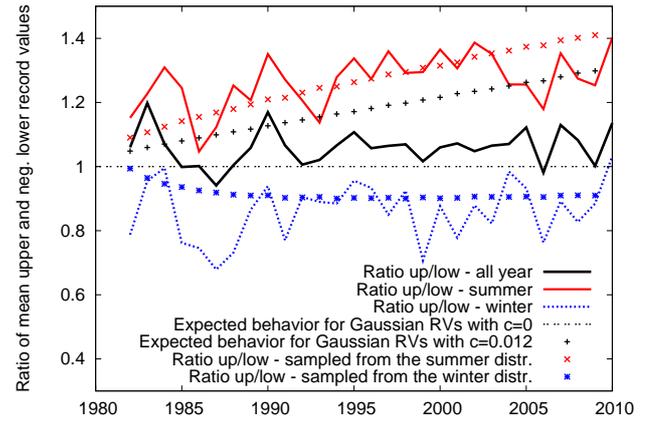}}
\caption{\label{Fig:Up_low_ratio} Ratio between rescaled upper and negative lower record values in EOBS2 for July and August (red line), January and February (blue line) and the entire calendar year (black line). The black crosses give the behavior of this ratio for a Gaussian LDM with a drift of $\tilde{c}=0.012\;y^{-1}$. The red and blue crosses give the development of the ratio for the summer and winter months when sampled from the respective observed distributions of daily temperatures in summer and winter. The thin black dashed line gives the behavior one would expect in the case of an iid Gaussian stationary climate. }
\end{figure}

With these findings it is clear how to explain the inconsistency of the top half of Fig. \ref{Fig:EOBS_mean_rec_at_n} with the analytical results for the LDM: A strong discrepancy of the behavior in the winter months averages out the effect of a slow positive increase on the record values in summer and leads to the fact that, when averaged over the entire calendar year, the upper and lower record values behave more or less in the same way.

To understand the anomaly in the winter months, we consider the seasonal variability of the pdf's of the daily temperature gridded values. In Fig. \ref{Fig:Norm_distr_winter_summer} we show the distributions of daily maximum temperatures in EOBS2 for two  months in summer (July and August) and two months in winter (January and February). The distributions were obtained after rescaling of the daily temperatures, as described above, so that they fitted a LDM with standard deviation one. The inset in Fig. \ref{Fig:Norm_distr_winter_summer} shows the analysis for the entire calendar year. Apparently the distribution for the winter months in data set EOBS2 is not Gaussian and has a much broader left tail. We believe that this asymmetry of the distribution is responsible for the anomalous behaviour of the mean record values in winter. 

\begin{figure}
\centerline{\includegraphics[width=0.48\textwidth]{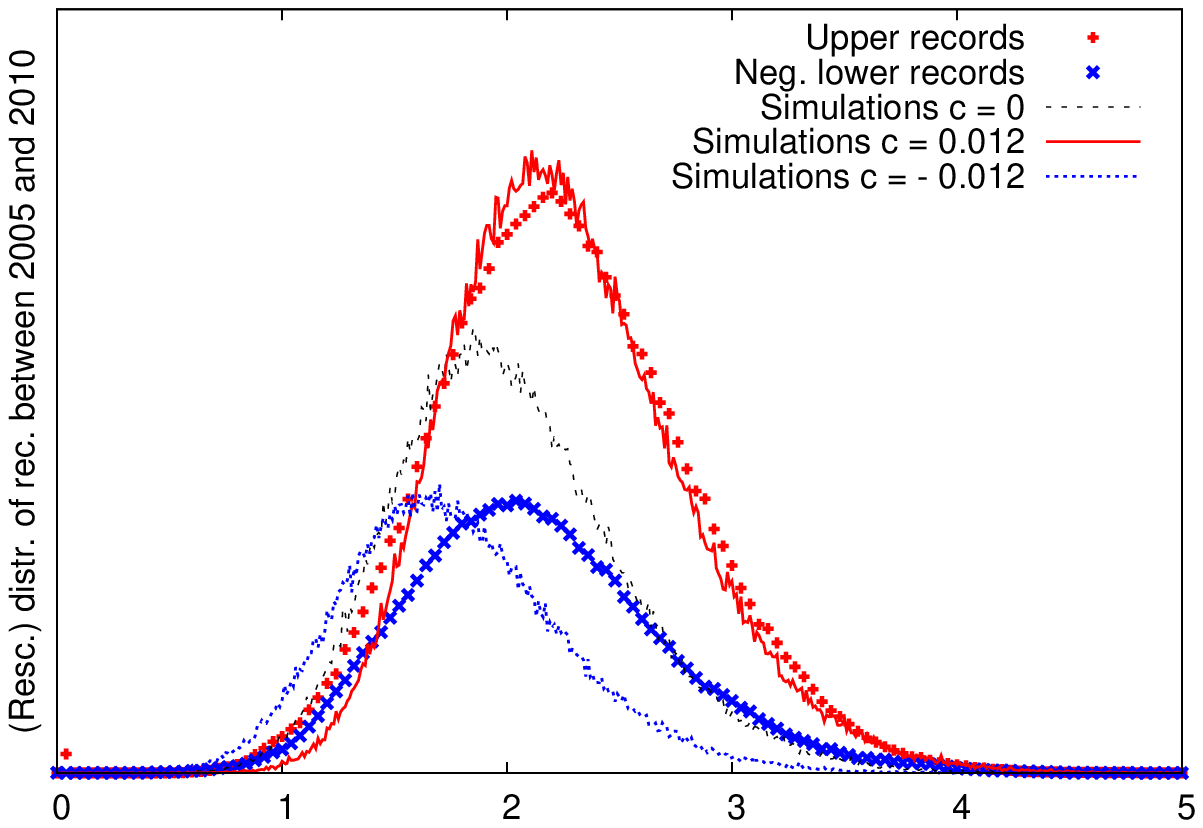}}
\centerline{\includegraphics[width=0.48\textwidth]{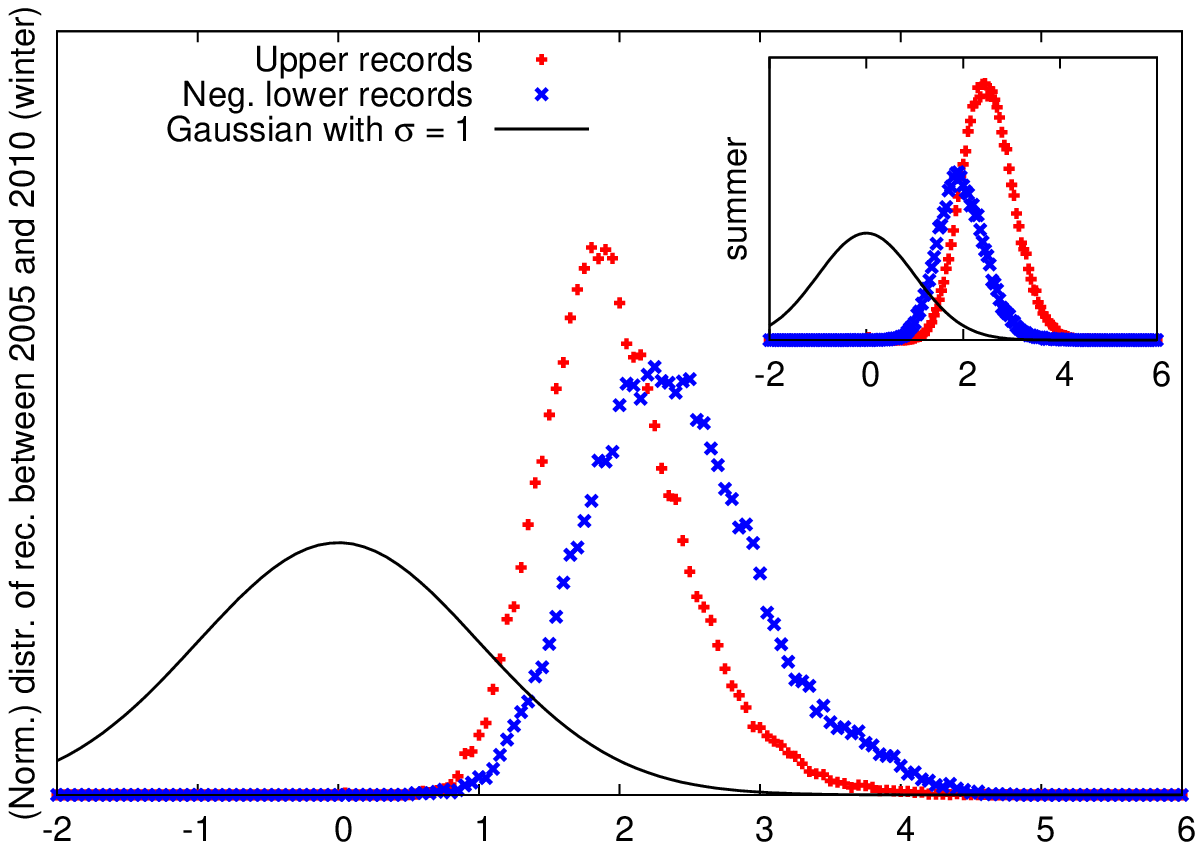}}
\caption{\label{Fig:Rec_at_5n_distribution} \textbf{Upper figure:}
  Estimated pdf's of record values in the rescaled temperature data
  from EOBS2 for the entire calendar year. The red crosses give the
  behavior of the upper records and the blue crosses the behavior of
  the negative cold records. The red and blue lines give the results
  from numerical simulations with a Gaussian LDM and a drift of
  $c=0.012$. The black line is the distribution one expects in a
  stationary climate ($c=0$) and Gaussian daily temperature
  measurements. All distributions are not normalized, the fact that
  the area under the curve from the upper records is much larger than
  the area under the corresponding curve from the lower record is a
  consequence of there being much more upper records than lower
  records. Note that, for these rescaled temperatures, the origin
  corresponds to the mean of daily temperatures in the initial
  year. \textbf{Lower figure:} The same analysis for the two
  considered months in winter (January and February) as well as, in
  the inset, the two months in summer (July and August). Numerical
  results are not given in this figure. For illustrative purposes we
  also plotted a Standard Normal distribution (with standard deviation
  $\sigma=1$), which in the framework of a Gaussian LDM, represents
  the rescaled distribution of daily temperatures in the first
  year. The area under the Gaussian equals the average of the areas under the shown probability densities of the upper and lower records. }
\end{figure}

In conclusion we find that while the ratio between rescaled upper and negative lower record values in summer is larger than one, it is less than one in winter, because of an asymmetric distribution of daily temperatures. Nevertheless, it might still be possible to describe the record values of the observations with a LDM, but we have shown that the underlying LDM for the winter months can not be based on a symmetric Gaussian distribution.

To further explore this asymmetry we performed numerical simulations based on the empirical pdf's of daily temperatures in EOBS2 for summer, winter and the entire year. In Fig. \ref{Fig:Up_low_ratio} we show the ratio between upper and lower record values in the data and compare them to the results from simulations with the distribution obtained from the data as well as the predictions from the Gaussian LDM. 

For the entire calendar year we find that the ratio remains close to one for the entire observation period, in contrast to a Gaussian LDM. However, if we consider the summer and winter months separately and compare the ratios obtained from the data with the ratios one obtains from a LDM with RV's sampled from the respective distributions of daily summer and winter temperatures, we find a good agreement with these non-Gaussian LDM's. In summer the ratio is strongly positive with upper records being generally further away from the increasing mean value than lower records. In winter, due to the skewness of the distribution, the situation is reversed and lower record values are shifting to be more extreme.

\subsection{Distribution of records in a certain year}

We also analyzed the full empirical pdf's of daily temperature records in the rescaled temperature data. Figure \ref{Fig:Rec_at_5n_distribution} shows the pdf's of temperature records for the entire calendar year (upper figure) and, again, the two considered months in winter and summer (lower figure) that occurred in the last five years of the observation period of EOBS2. We did not normalize the distributions for illustrative purposes. In the upper figure we compared the observational distributions with numerical results from the Gaussian LDM with a normalized drift of $\tilde{c}=0.012\;y^{-1}$.

We find that while the rescaled upper record values are in good
agreement with the Gaussian LDM, the distribution of the lower record
value is significantly broader than expected from the simulations. The
pdf's of the upper and lower record values seem to have more or less
the same mean value, which is in good agreement with our findings in
the previous section and Fig. \ref{Fig:EOBS_mean_rec_at_n}. The pdf's
in this figure are not normalized, the total area under the curves
corresponds to the number of records that occurred in the last five
years. As a result the upper figure shows that there were many more upper records than lower records in the data, but the shapes of the pdf's of upper and lower records look very similar.

\begin{figure}
\centerline{\includegraphics[width=0.48\textwidth]{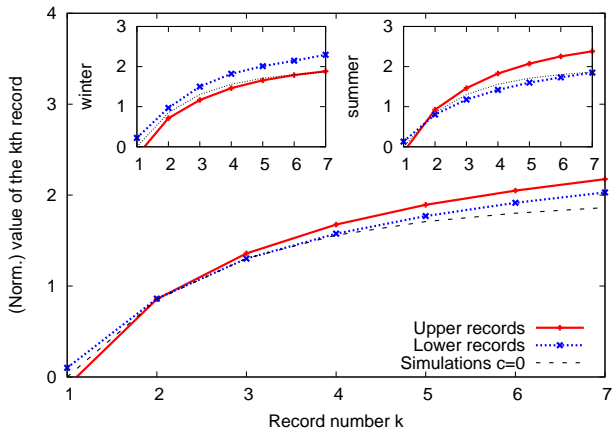}}
\caption{\label{Fig:Mean_kth_rec} Mean values of record-breaking temperatures with a fixed record number $k$ in EOBS2 data using the same rescaling as for the record values with fixed time of occurrence $n$. The red lines show the time series of the upper record values depending on $k$, the blue lines the corresponding series of the lower records. The black dashed line shows the growth of the mean of records with record number $k$ for Gaussian RV's without drift. The insets show the same analysis for January and February (left inset) and July and August (right inset).}
\end{figure}

The pdf's in the lower figure are also in good agreement with the mean values in Fig. \ref{Fig:EOBS_mean_rec_at_n}. The pdf's for the winter months show that in winter the lower records were more extreme with a mean value further away from the mean of the daily maximum temperatures.

Also the width of the pdf of the lower records is larger than in the case of the upper records, so the values of lower records have a large inherent uncertainty. While there are almost no upper records more than $3\sigma$ away from the average behavior, there is a large number of lower records that exceeded this barrier, some of them were even beyond $4\sigma$, which, given the large standard deviation of daily temperatures in winter (up to $8\;K^{\circ}$), is a huge fluctuation. The inset in the lower figure gives the same plot for the summer. Here, in agreement with a Gaussian LDM, upper records are more extreme than lower records.

\subsection{Records with a given record number}

We also analyzed the statistics of record temperatures in the data set EOBS2 with a given record number $k$. For this purpose we performed the same rescaling as described above in the context of records with a fixed time of occurrence $n$. The results of the analysis for two months in winter, in summer and over the entire calendar year in data set EOBS2 can be found in Fig. \ref{Fig:Mean_kth_rec}. The figure shows the estimated mean value of records in the time series of the rescaled entries $\tilde{T}_i$ (Eq. (\ref{rescaling})) plotted against the record number $k$. 

This analysis of the mean record value with fixed record number is in good agreement to our findings in the above analysis of the mean record values with fixed occurrence time $n$. The analysis for the entire calendar year shows that the behaviour of the mean is again comparable for the upper and lower records. Similar results were already obtained by Redner and Peterson \citeyearpar{Redner2006}, who studied the mean values for the $k$th records in the station data from Philadelphia. In this study, the (rescaled) upper and lower record values for small values of $k<6$ were almost identical as well.

If we consider the two winter months January and February (left inset), we find again that cold records are further away from the mean value than the upper records and are therefore more extreme. In July and August (right inset) it is exactly the other way around. Here the LDM based on symmetric, Gaussian RV's works very well and mean record values show the expected asymmetry with more extreme upper records due to the positive trend.

\begin{figure}[t]
\centerline{\includegraphics[width=0.48\textwidth]{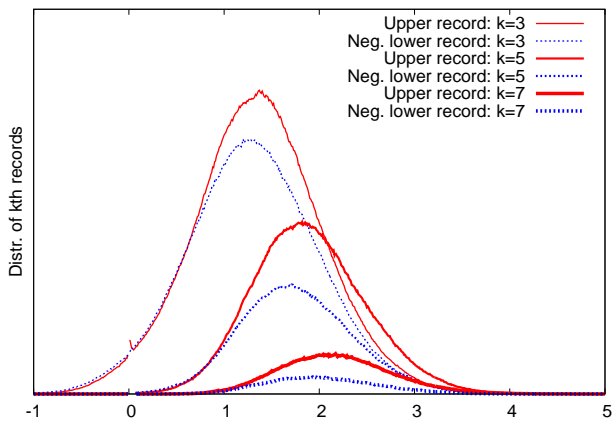}}
\caption{\label{Fig:Distr_kth_recs} Estimated pdf's of record values with a given record number $k$ of the rescaled daily maximum temperatures from EOBS2. The red lines give the pdf's of upper records with record numbers $k=3,5$ and $7$. The blue lines give the pdf's of the negative lower records with the same record numbers $k$. Note that, as in Fig. \ref{Fig:Rec_at_5n_distribution}, the density functions are not normalized.}
\end{figure}

In Fig. \ref{Fig:Distr_kth_recs} we also show the estimated pdf's of record values for some selected record numbers $k$ and the entire calendar year. Again, the density functions are not normalized, so the area under the curves corresponds to the total number of records with a given record number. Apparently, even though in all cases the number of upper records was higher than the number of lower records, the shapes of the pdf's look very similar. 

\subsection{Record values in the far north}

A reason for the asymmetry of the daily temperature distribution in winter may be that some parts of Europe, especially in the north, are covered by snow during the winter months. Snow cover decouples the atmosphere from the soil through its isolating effects. Further a snow surface is a very efficient black body radiator. Therefore snow covered surfaces tend to produce very low near surface atmospheric temperatures especially under conditions of small insolation \citep{Geiger1995}. 

\begin{figure}[t]
\centerline{\includegraphics[width=0.48\textwidth]{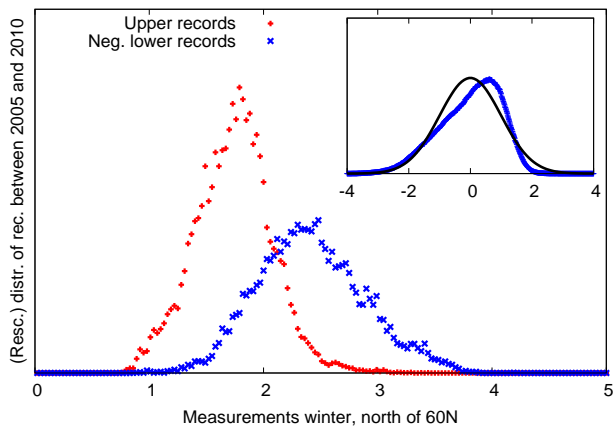}}
\caption{\label{Fig:Far_north} Estimated pdf of record values in the rescaled temperature data from EOBS2 for the months of January and February in the regions north of $60^{\circ}N$. The red crosses give the behaviour of the upper records and the blue crosses the behaviour of the lower records. The inset shows the estimated pdf of rescaled daily temperatures in these two months and in the region north of $60^{\circ}N$. Note that, as in Fig. \ref{Fig:Rec_at_5n_distribution}, the density functions are not normalized.}
\end{figure}

In Fig. \ref{Fig:Far_north} we show the pdf's of records that occurred in the last five years of the EOBS2 data only for grid points north of $60^{\circ}N$. Apparently, here the effect of the asymmetry in the daily temperature pdf's on the record values is even stronger. The estimated density functions of the lower records is centered more than one standard deviation further away from the mean than the density functions of the upper records. The inset shows the pdf of the daily temperatures for this region, which is much more heavily skewed than the pdf for the entire data set of EOBS2 (compare to Fig. \ref{Fig:Norm_distr_winter_summer}).

\section{Discussion \& Conclusion}

In this article we analyzed both the probability of occurrence and the
probability density function of record breaking temperature values in
Europe and the United States. After presenting some analytical
findings on the Linear Drift Model (LDM) in section 2, we analyzed the
occurrence of records in daily and monthly data in sections 3 and
4. In agreement with earlier work by \cite{Wergen2010} we found a
significant effect of the observed slow positive trend of temperature
on the number of upper and lower records. This effect can be described
by the LDM up to some accuracy. The effect of increasing temperatures
on the monthly mean temperatures is clearly stronger than the effect
on the daily measurements because of the smaller standard deviation of
those averages. During certain seasons in the United States we found
up to nine times more upper records than lower ones at the end of the
51 years of observation 1960-2010. This also explains the findings of
Rahmstorf and Coumou \citeyearpar{Rahmstorf2011}, as well as Coumou et al. \citeyearpar{Coumou2013}, who studied monthly
and annual mean temperatures and found the strongest effect of warming
for the annual global-mean temperature. The global mean has a very
small standard deviation of $\sigma=0.09\;\textrm{K}^{\circ}$ \citep{Rahmstorf2011} which
implies that already a very small drift can measurably increase the upper record rate. 

In section 5 we presented new results on the mean value of a record that occurs at a certain time in the LDM. We were able to obtain analytical results of the effect of a linear drift on the mean value of a record in case of Gaussian RV's in the important regime of $cn\ll\sigma$. In section 6 we compared these results to the gridded data set EOBS2. When analyzing the entire calendar year, a Gaussian LDM fails to describe the effect of slowly increasing temperatures on the mean record values. The reason for this failure is a pronounced asymmetry of the daily temperature probability density function in winter. In contrast the record values during the summer months can be well described by a Gaussian LDM.
%, an effect which is known to meteorologist and originates partly from the high albedo of snow (\textbf{citations needed}). 

When we use a non-Gaussian LDM with RV's sampled from the daily temperature probability density of the considered winter months, this model is also capable of describing temperature record values in winter. As a general result the lower records in winter are significantly more extreme than upper records because of this asymmetry. This leads to the unintuitive results that lower records of near surface temperature in the case of slowly increasing temperatures occur with a reduced probability but once they occur they are more extreme than their upper counterpart due to the pronounced asymmetry of the daily temperature values in winter. When we consider grid points in the northern parts of Europe this effect is even stronger. Here lower records are on average more than one standard deviation further away from the mean than upper records and the estimated density of lower records is also clearly broader than that of the upper records.

It might be interesting to further explore the effect of specific weather conditions especially in northern Europe, but also in other regions, on the statistics of record values. In particular it remains an open question, how strongly the occurrence of very extreme lower records in winter correlates with snow coverage and other meteorological events, such as winter blocking highs. It is however clear that the fact, that we still encounter quite extreme cold streaks in winter, particularly in northern regions, is not in contradiction with global warming and only a consequence of the very skew distribution of daily temperatures in winter.

\begin{acknowledgements}
GW is grateful for financial support provided by DFG within the Bonn Cologne Graduate School of Physics and Astronomy as well as the Friedrich-Ebert-Stiftung. We acknowledge the E-OBS dataset from the EU-FP6 project ENSEMBLES and the data providers in the ECA\&D project.
\end{acknowledgements}

\bibliographystyle{spbasic}      % basic style, author-year citations
%\bibliographystyle{paper-en}
%\bibliography{literature}   % name your BibTeX data base

\end{document}